\documentclass[sigconf]{acmart}

\AtBeginDocument{%
  \providecommand\BibTeX{{%
    \normalfont B\kern-0.5em{\scshape i\kern-0.25em b}\kern-0.8em\TeX}}}

\copyrightyear{2023}
\acmYear{2023}
\setcopyright{rightsretained} 
\acmConference[SIGCSE 2023]{Proceedings of the 54th ACM Technical
Symposium on Computing Science Education V. 1}{March 15--18,
2023}{Toronto, ON, Canada}
\acmBooktitle{Proceedings of the 54th ACM Technical Symposium on
Computing Science Education V. 1 (SIGCSE 2023), March 15--18, 2023,
Toronto, ON, Canada}
\acmDOI{10.1145/3545945.3569855}
\acmISBN{978-1-4503-9431-4/23/03}

\hypersetup{
       breaklinks=true,   
       colorlinks=true,   
    }

\makeatletter
\gdef\@copyrightpermission{
  \begin{minipage}{0.3\columnwidth}
     \href{https://creativecommons.org/licenses/by/4.0/}{\includegraphics[width=0.90\textwidth]{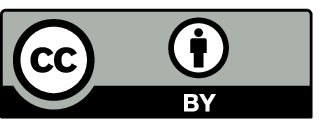}}
  \end{minipage}\hfill
  \begin{minipage}{0.7\columnwidth}
     \href{https://creativecommons.org/licenses/by/4.0/}
     {This work is licensed under a Creative Commons Attribution International 4.0 License.}
  \end{minipage}
  \vspace{5pt}
}
\makeatother

\usepackage{multirow}
\usepackage{balance} 

\begin{document}

\title[Incorporating Ethics in Computing Courses: Perspectives from Educators]{Incorporating Ethics in Computing Courses: \linebreak Barriers, Support, and Perspectives from Educators}

\author{Jessie J. Smith}
\email{jessie.smith-1@colorado.edu}
\affiliation{%
  \institution{University of Colorado, Boulder}
    \country{Colorado, USA}
}
\author{Blakeley H. Payne}
\email{blakeley.hoffman@colorado.edu}
\affiliation{%
  \institution{University of Colorado, Boulder}
    \country{Colorado, USA}
}
\author{Shamika Klassen}
\email{shamika.klassen@colorado.edu}
\affiliation{%
  \institution{University of Colorado, Boulder}
    \country{Colorado, USA}
}
\author{Dylan Thomas Doyle}
\email{dylan.doyle@colorado.edu}
\affiliation{%
  \institution{University of Colorado, Boulder}
    \country{Colorado, USA}
}
\author{Casey Fiesler}
\email{casey.fiesler@colorado.edu }
\affiliation{%
  \institution{University of Colorado, Boulder} 
  \country{Colorado, USA}
}

\renewcommand{\shortauthors}{Jessie J. Smith, Blakeley H. Payne, Shamika Klassen, Dylan Thomas Doyle, \& Casey Fiesler}

\begin{abstract}
Incorporating ethics into computing education has become a priority for the SIGCSE community. Many computing departments and educators have contributed to this endeavor by creating standalone computing ethics courses or integrating ethics modules and discussions into preexisting curricula. In this study, we hope to support this effort by reporting on computing educators' attitudes toward including ethics in their computing classroom, with a special focus on the structures that hinder or help this endeavor. We surveyed 138 higher education computing instructors to understand their attitudes toward including ethics in their classes, what barriers might be preventing them from doing so, and which structures best support them. We found that even though instructors were generally positive about ethics as a component of computing education, there are specific barriers preventing ethics from being included in some computing courses. In this work, we explore how to alleviate these barriers and outline support structures that could encourage further integration of ethics and computing in higher education.
\end{abstract}

\ccsdesc[500]{Social and professional topics~Computing education}



\keywords{Ethics, Post-Secondary, Curriculum, Perspectives, Survey}

\maketitle

\section{Introduction \& Background}

"How can we support ethics education in computer science?" is an ongoing, important topic in the SIGCSE community and related communities \cite{Quinn2006-QUIOTC, fiesler2020we}. Researchers and educators have sought to address this question through a variety of means, such as creating open-sourced lessons and materials, reporting on in-classroom experiences, making arguments about the importance of computing ethics education, and discussing what such an education might prioritize. Many of these efforts address specific challenges computing educators face when integrating ethics education into computing curricula. For example, \citet{10.1145/3478431.3499291} addresses the political issue of getting computing faculty to see ethics as an integral part of the classroom. Another common challenge often cited in the literature is around faculty expertise and time. \citet{fiesler2021integrating} name the time constraints often placed on instructors both in and outside of the classroom when discussing incorporating ethics content into introductory programming classes. One solution to this is the creation and sharing of open-sourced classroom activities which either focus on an ethical topic or provide ways to infuse lessons on traditional computing topics with ethics \cite{safest-path, KlassenFiesler2022, cs-ethics,doore2020assignments}. A number of repositories have also been created, which include resources such as syllabi, reading lists, or curricular modules \cite{fiesler2020we, responsible-problem-solving}. 

Another solution offered to alleviate lack of faculty expertise or lack of time is the inclusion of undergraduate and graduate students as ethical curriculum developers and educators. The role students play in ethics education can be seen in both standalone courses \cite{Teaching-Ethics-Pedagogy}, as well as in departmental or university-level efforts such as the Embedded Ethics program at Harvard University, or the Socially Responsible Computing program at Brown University \cite{Embedded-EthiCS,Brown-Model}. However, teaching assistants also report time constraints as a challenge in the classroom \cite{ta-challenges}. 

Other work has addressed pedagogical challenges for incorporating ethics in computing courses by introducing activities such as mock trials or writing about current events, which can be scaled to work for different ethical topics \cite{Mock-Trial,learner-centered,role-play}. \citet{grading} have offered practical guidance for grading computing ethics assignments, which alleviates the challenge that ethics assignments are perceived as having ``no correct answers.''

Though this previous work is promising for alleviating some of the barriers to incorporating ethics in the computing classroom, computing instructors still face a number of challenges when seeking to integrate ethics content into their courses. In this work, we expand on previous research by identifying and characterizing the primary challenges educators still face when including ethics in their computing classrooms, as well as the resources and structures which enable them to be successful. To do this, we center the voices of instructors by probing their perceptions and experiences with ethics in computing education. In this study, we surveyed 138 higher education computing instructors, focusing on the following research questions:


\begin{itemize}
    \item \textbf{RQ1:} What perceptions do computing professors have about ethics education in the computing classroom?
    \item \textbf{RQ2:} What barriers prevent computing professors from incorporating ethics in their computing classroom?
    \item \textbf{RQ3:} What support do computing professors need to help them incorporate ethics in their computing classroom?
\end{itemize}

We present our findings from this survey below, and then discuss the implications of the results. We offer suggestions for overcoming and thinking about common barriers computing educators face, as well as recommendations for bolstering support.

\section{Methods}
For this study, we recruited participants to complete a survey about their perspectives and experiences with integrating ethics into computing courses they had previously taught\footnote{The authors are happy to provide survey materials upon request.}. It is important to note that ``ethics'' is a broad concept in the context of computing and computing education, and often serves as shorthand for a collection of concepts such as responsibility, social impact, and justice \cite{fiesler2020we}. For example, the accrediting body for computer science programs in the U.S. requires students to have ``an understanding of professional, ethical, legal, security and social issues and responsibilities,'' a requirement that departments interpret in a variety of ways \cite{homkes2009meeting}. One might consider ethics as being about doing ``the right thing'' \cite{aiken1983reflections},  but at the same time there are often not ``right'' and ``wrong'' answers and ethics pedagogy frequently focuses on critical thinking or issue spotting \cite{fiesler2020we}.  Educators are also increasingly considering their role in making injustices visible for students \cite{ko2020time}, and Ferreira and Vardi point to social justice as the ``most important issue confronting computer science today'' and critical to ``deep'' tech ethics \cite{ferreira2021deep}. With this in mind, during the survey, we did not prompt participants with a concrete definition of ethics in order to assess their perceptions based on their personal definitions and experiences with how this collection of concepts under an umbrella of ethics fit into computing education.

At the beginning of the survey, we asked participants to choose one computing course they had taught most often in the last two years and/or the course they thought was most representative of their teaching. About half of the questions in the survey were focused on this specific course, while the remaining questions were focused on the participants' general attitudes towards including ethics in computing education more broadly.


We recruited educators who were currently teaching or had taught computing courses at a higher-education\footnote{\textbf{Higher-education} for the context of this survey includes university, college, post-secondary, community college, and any education beyond high school.} institution in the past two years. We recruited through social media and targeted emails. Emails used in recruitment were open-sourced from \citet{barker2015influences}, who originally downloaded, cleaned, and categorized emails from lists created by the Integrated Postsecondary Education Data System (IPEDS) \cite{ipeds}. The recruitment text described the survey's goals as: \emph{``to better understand your thoughts towards including ethics content in your curriculum.''} Although we specified in our recruitment material that we were interested in a variety of experiences (including educators who have never considered including ethics in their course or who feel strongly about excluding ethics), there may still have been selection bias towards educators who were interested in ethics, and thus the results of this study may not reflect the overall population of computing educators. Participation in the survey was voluntary and uncompensated.


A total of 138 participants completed the full survey. The United States of America represented 91.7\% of participants, and the following countries represented less than 2\% of respondents each: Australia, Canada, France, Germany, India, Mexico, Netherlands, and the United Kingdom. The full demographic breakdown of these participants is shown in Table \ref{table:participant-demographics}. Each participant's responses were labeled with a unique participant id number (e.g., P1 or P50), which we reference throughout Section \ref{findings}.

\begin{table}[]

 \caption{Demographic breakdown of survey respondents. Participants who preferred to not disclose or did not answer demographic questions were left out of this table. For the ``\% of Total'' column, n=138.}
 \label{table:participant-demographics}
\begin{tabular}{p{1.2cm}|p{2.9cm}|p{0.9cm}|p{1.35cm}}
\textbf{Category}& \textbf{Demographic}& \textbf{Raw \#} & \textbf{\% of Total} \\

\hline
\multirow{2}{1.2cm}{Course Type} & Included Ethics                          & 76                                  & 64.4\%                                                  \\
                                         & Excluded Ethics                   & 42                                  & 35.6\%                                                  \\
\hline
\multirow{4}{1.2cm}{Gender}                  & Woman                                    & 39                                  & 34.5\%                                                  \\
                                         & Man                                      & 67                                  & 59.3\%                                                  \\
                                         & Non-Binary                               & 2                                   & 1.8\%                                                                                      \\
\hline
\multirow{6}{1.2cm}{Age}                     & 25-34                                    & 20                                  & 17.7\%                                                  \\
                                         & 35-44                                    & 35                                  & 31\%                                                     \\
                                         & 45-54                                    & 23                                  & 20.4\%                                                  \\
                                         & 55-64                                    & 18                                  & 15.9\%                                                  \\
                                         & 65-74                                    & 12                                  & 10.6\%                                    
                                                   \\
\hline
\multirow{7}{1.2cm}{Role of Educator}        & Tenured/Associate/Full                        &64                                  & 46.4\%                                                  \\
                                         & Assistant/Untenured                      & 18                                  & 13\%                                                  \\
                                         & Instructor                      & 11                                  & 7.9\%                                                  \\
                                         & Adjunct & 8                                  & 5.8\% \\
                                         & Lecturer & 9                                  & 6.5\% \\
                                         & Student Instructor & 10                                  & 7.2\% \\
                                         & Emeritus & 1                                  & 0.7\%                                                             \\
\hline
\multirow{7}{1.2cm}{Race and/or Ethnicity}   & White/Caucasian                 & 85                                  & 78.7\%                                                  \\
                                         & Mixed                                    & 5                                   & 4.6\%                                                   \\
                                         & South Asian                              & 5                                   & 4.6\%                                     
                                                 \\
                                         & Asian                                    & 3                                   & 2.8\%                                                   \\
                                         & Black                                    & 3                                   & 2.8\%                                                   \\
                                         & Hispanic                                 & 1                                   & 0.9\%                                                   \\
                                         & Middle Eastern                           & 1                                   & 0.9\%
\end{tabular}
\end{table}

The survey data was analyzed using both quantitative and qualitative methods. For quantitative methods, we generated descriptive statistics and performed Chi-squared tests to determine statistically significant relationships between survey answers and respondent demographics. For qualitative methods, we implemented a qualitative coding approach informed by thematic and content analysis to categorize themes from free-form participant responses. One member of the research team coded themes from written responses, while the remaining members checked the themes and categories until agreement was reached.


\section{Findings}
\label{findings}
We organize our findings around each of our research questions. We begin by describing participants' general attitudes towards ethics in the computing classroom (RQ1). We then explore the barriers that hinder participants from including ethics in their computing courses (RQ2), and conclude by describing the structures and support that might be helpful for participants to begin or continue including ethics in their computing courses (RQ3).

\subsection{RQ1: Educators' Perceptions of Ethics}
In general, the computing educators who took this survey expressed positive perceptions of ethics in the computing classroom. 71.5\% of participants ``disagreed'' or ``strongly disagreed'' that ethics topics should be taught by faculty from departments like Philosophy or Sociology rather than by computing faculty. Additionally, 61.3\% of respondents ``agreed'' or ``strongly agreed'' that their colleagues think including ethics in the computing classroom is important. 93\% of participants ``agreed'' or ``strongly agreed'' that including ethics in the computing classroom prepares students for a career in computing after graduation, and 93\% ``agreed'' or ``strongly agreed'' that students gain value from discussing ethics in the computing classroom. 76\% of respondents reported that ethics does not take away from a student's ability to learn core computing topics. Previous work examining student attitudes towards ethics content in their computing coursework has also suggested their perception of its value \cite{fiesler2021integrating, KlassenFiesler2022}. However, a recent study suggested that though students agreed their ethics education might eventually help them in the workplace, it would not help them in obtaining a job \cite{sarder-computing}.


Despite the overall support for ethics in the general sense, 51\% of respondents either ``agreed'' or ``strongly agreed'' that some topics in computing do not have an ethical consideration or component--this included both respondents who included ethics in their course and those who did not. For example, multiple participants who did not include ethics in their courses noted that these courses focused on mathematical concepts and thus did not apply to ethics topics; for example, P30 elaborated: \emph{``It's a math class about computing.  There are not a lot of ethical issues.''} However, P89 (who included ethics in their course), had the opposite opinion about a similar class:

\begin{quote}\emph{``Someone I admire in computing education wrote earlier this year that he thought it was more important that his computing students understand communities and context than details of the mathematics underlying computing... so I thought about what it would look like to think about communities and context within discrete mathematics/discrete structures courses. And it does feel like a disservice to teach anything touching on algorithms without talking about the conversations around algorithmic bias right now,''} (P89).
\end{quote}

Participants also expressed mixed responses about \emph{when} in the computing degree it is most appropriate to incorporate ethics. 23\% of participants reported that they believe students need to have a baseline understanding of computing skills and topics before they can think about ethical issues; 62\% of participants who reported this opinion did not include ethics in their own course.

In summary, most survey respondents saw value in ethics integration, but may have felt that it was not appropriate for their specific course. Our results also uncover additional barriers that may be preventing educators from including ethics in their computing classroom.

\subsection{RQ2: Barriers to Ethics Integration}
42 out of 138 participants did not include ethics in the course they chose to explore in the survey. For these 42 participants, we were curious if there were any barriers that hindered their ability to include ethics in their course. Of those 42 participants, 40\% reported that specific barriers were preventing them from including ethics in their course, while 60\% reported that there were no specific barriers. For all of these participants, regardless of whether they self-reported if they experienced ``barriers,'' we asked them to describe the reasons why they chose to exclude ethics from their course.

We categorized these reasons into six main themes, which we describe through participant quotes and responses below. Individual quotes were selected as representatives of the broader themes.
\begin{enumerate}
    \item \textbf{The desire to leave ethics to other courses or other departments.} This desire was described by P95 who said, \emph{``I am not an expert on ethics the way that I am an expert on algorithms, and so do not feel comfortable teaching content that I am not familiar with as part of my research focus.  I would like to defer to the philosophy department, who have actual expertise in moral philosophy.''}
    \item \textbf{Other topics having higher priority.} This prioritization was described by P8 who said, \emph{``[there is] too much technical material to cover; ethics covered elsewhere in CS curriculum.''}, or P24 who said, \emph{``ethics may be relevant to the course, but it is far less relevant than the course content, so I leave it out.''}, or P59 who said \emph{``there is so much technical content that I need to cover that I often just focus on that.''}
    \item \textbf{The feeling that course content is not related to ethics.} P88 described this feeling by sharing that, \emph{``[I teach] an algorithms course. As passionate as I am about ethics, it really doesn't fit in.''} P88 also shared that there were likely ethical components related to their course topic, but they had difficulty knowing what they were. \emph{``I think it's really unlikely that there are topics in computer science that have absolutely no relevant ethical discussion. I just haven't thought of ethical issues that are relevant to the topic of my class. If I were teaching machine learning or something where the connection is obvious I'd definitely already be incorporating ethics.''} (P88).
    \item \textbf{The inability to control the course curriculum.} This theme was described well by P22 who said, \emph{``[ethics] is not part of the design of the course. I am not on the curriculum committee and so can’t change this.''} Another participant (P12) similarly described that they did not include ethics because it was not in the textbook that their course was based on, which they had no control over.
    \item \textbf{Not having enough time, incentives, or encouragement.} Though our participants had mixed responses about the time commitment needed to include ethics in their course, 68.1\% of participants ``agreed'' or ``strongly agreed'' that including ethics requires a substantial time commitment from educators. Several participants mentioned that in order to alleviate this challenge, they require more time to create ethics material, more time to teach it to students, and greater incentives to spend the time doing these things. This desire was summed up nicely by P42 who said, \emph{“I may need to remove a topic or two that I currently teach in order to have time for ethics discussions. I would also need to potentially adjust my grading system, if I added ethics assessments. Time and encouragement are both lacking,”} (P42).
    \item \textbf{The feeling that ethics distracts students from core computing concepts.} Within the group of respondents who do not include ethics in their courses, 40.4\% ``agreed'' or ``strongly agreed'' that ethics takes away students' attention from core computing concepts. For example, P53 wrote, \emph{``ethics instruction is unquestionably essential for computer science students. Ethical mistakes in computing can have disastrous consequences. However, adding ethics to courses on algorithms or discrete mathematics is distracting and ineffective.''} This perspective has been challenged in prior work, such as a recent paper from \citet{icer-brown} that provides resources for introducing ethics into an algorithms class in a way that both addresses technical and ethics knowledge.
    
\end{enumerate}

\subsubsection{\textbf{Confidence and Competence.}} One respondent mentioned that the barrier preventing them from including ethics in their course was that they were \emph{``not sure how to link [ethics] to the [computing] material and how to develop the skills to do this,''} (P70). Given that computing educators might not be trained in disciplines like ethics, we were curious if a lack of confidence for teaching ethics content might be a barrier to incorporating ethics discussions in their course. We discovered that regardless of whether survey respondents included ethics in their course or not, and regardless of whether ethics was required to be included in their course or not, on average all respondents felt moderately confident about incorporating ethics. 67\% of all respondents stated they either “agreed” or “strongly agreed” that they felt confident in their ability to include ethics in the computing classroom. Only 13\% of all respondents “disagreed”, and 2\% of all respondents “strongly disagreed” with that statement. This suggests that among our respondents, lack of confidence was not a significant barrier to including ethics in their computing courses. However, despite participants' confidence in including ethics, some expressed concern that incorporating ethics in the computing classroom would not be helpful if instructors do not have the appropriate background to do so, a concern that has also been addressed in prior work \cite{KlassenFiesler2022}. For example, P118 wrote: 

\begin{quote}
\emph{``I'm not confident that many people teaching these classes are competent to teach on ethical issues. Like... It's not in their work or their training so far, not part of the regular reading, so why trust them to teach it? To be clear I think it belongs in their classes, I just worry they'll do it poorly,''} (P118).
\end{quote} 

This finding suggests that computing educators might benefit from additional training or support structures to avoid incorporating ethics ``poorly.''


\subsection{RQ3: Structures Supporting Ethics}
Here we describe some of the structures that support incorporation of ethics in computing courses, as reported or observed by participant responses. Of the 76 survey respondents who included ethics in their course, we asked them which resources they used when incorporating ethics in their classroom, and which of those resources was the single most helpful for them. The top two most helpful resources for these survey respondents were (1) online resources (reported by 55\% of these select participants); and (2) collaboration with others within their institution (reported by 33.4\% of these select participants). We also conducted statistical testing to observe any relationships between all participants' responses that might suggest additional support structures. We find that regardless of whether participants included ethics in their course, the three most prominent themes for support structures were:
\begin{enumerate}
    
    \item Resources (especially found online) that have already been created or resources for generating material from scratch.
    \item Collaboration, encouragement and support from colleagues within the institution to incorporate ethics into the computing classroom.
    \item Having a professional community to discuss ethics with.
\end{enumerate}

\subsubsection{\textbf{Additional Resources.}} The most helpful support structure reported by participants who included ethics in their course was online resources (e.g., open-source coding assignments that incorporate ethics, or reading lists found online). This finding tracks to prior work in which educators were interviewed about using a particular ethics-related teaching exercise. In this prior work, many educators reported that they used this exercise because they found it online and it was easy to adapt from the provided materials \cite{KlassenFiesler2022}. Similarly, in our data, one respondent reported that online resources were the most helpful because they can be applicable to current events, and it \emph{“helps [them] stay up to date with situations in the news,”} (P20). P23 agreed with this advantage, especially with respect to social media:

\begin{quote}\emph{“I pick up a lot from Twitter, actually, amongst other sources. So I'm constantly aware of what seems to be going on. As a result, when we made a department-wide effort to add this material, I didn't have to scramble; for each assignment, natural content just fell out, and my only embarrassment was not adding it sooner,”} (P23).
\end{quote}

Although online resources seemed to be helpful for many introductory computing topics, as mentioned by P65 – \emph{“there are good introductory resources available online that introduce students in "CS 0" to topics such as algorithmic bias, the digital divide, ethical hacking, underrepresentation in computing, and more,”} – there appeared to be a dearth of content about more niche or advanced computing topics. This led some educators to create their own content. Some educators saw this as a benefit, such as P6 who said that starting from scratch \emph{“allowed me to make most use of my experiences and still fit into the course,” }(P6).

However, for others, creating content from scratch was a necessity, pointing again to the need for more resources for higher level and niche computing classes. For example, P41 wrote, \emph{“there is very little existing material that focused on the intersection of ethics and the course topic. Plenty of material on more general ethics, but little of it is specifically about the course's topic, which I wanted to emphasize,”} (P41). And even among our respondents who made the effort to integrate ethics, many noted that it could be quite time consuming, such as P11 who stated, \emph{“finding good materials takes time. Since I have been teaching for 20+ years, I have developed good material (over time) that inspires classroom discussion.”} Other respondents mentioned that they created all ethics content in their course from scratch out of necessity, because they were given no other resources from their university. \emph{“The course was demanded as a standalone course to make the degree program look more aligned to the mission-centric board of directors. Admin provided no guidance or support for course content, merely a course number and title,”} (P17).  In summary, whether participants turned toward creating their own material from scratch or not, providing additional resources (online or otherwise) appeared to be a helpful and necessary support.

\subsubsection{\textbf{Collaboration with Others.}} The second most helpful resource selected by participants who included ethics in their course was collaboration with others at their university (e.g., faculty or grad students), which was chosen by 32.6\% of these selected participants. P69 shared that they found collaborating with other faculty helpful for incorporating ethics in their course because \emph{“each faculty member can bring their sources to the table so that all of the collaborators have more places to look for new learning opportunities about ethics + computing. We also get to hear reactions to our early thinking and shape messages together.”} P86 agreed with this sentiment: \emph{``a lot of the online resources I looked at just weren't that helpful for what we were looking for. Talking with other faculty was much moreso.''} P133 also included that collaborating with others to create the ethics content for their course gave them \emph{``access to multidisciplinary perspectives on content and pedagogical strategies.''} Though collaborating with faculty was helpful for some, one respondent mentioned that collaborating with faculty was not useful for them because they \emph{“teach at a remote school… where people have questionable ethics.”} For this participant, \emph{“finding external sources online is the most functional and useful,”} (P25). This finding could imply that faculty act as both a support structure and an obstacle for including ethics; with faculty who support ethics acting as the former, and those who do not support ethics acting as the latter.

\subsubsection{\textbf{Community.}} During our data analysis, we compared participant's survey responses (excluding demographic questions) between those who included ethics in their course and those who excluded ethics in their course to explore potential relationships between these variables. We found one statistically significant relationship: the relationship between participants who incorporated ethics in their courses and participants who had a professional community to discuss ethics with. When educators had access to community support, they were more likely to incorporate ethics into their courses, and visa versa -- without community support, they were more likely to exclude ethics from their course. This relationship was observed through the Chi-squared test ($\chi^2$ = 16.4, p < 0.01) with a Cramér's V effect size of 0.376. The communities that participants discussed ethics with included their department, peers, research community, conference community, Twitter/other online communities, or other. The distribution of responses was even between these options. Of those who selected “other,” their communities included: university, service, mailing list, and faith community. For participants who included ethics in their course, we asked if there were any additional resources or assistance that would help them continue to integrate ethics in their course and/or to do so more successfully. Several participants mentioned community in their response, such as P101 who said that it would help them to have \emph{``a stronger professional community around this topic.''} 



\section{Discussion}
This study provides opportunities for the SIGCSE community to explore resources that best support integration of ethics into computing courses. Supporting ethics integration may require shifts at the classroom, department, and institutional levels. In this section we discuss what some of these interventions could look like, and provide recommendations for future work that could address barriers while also strengthening support structures.

\emph{\textbf{Cultivating Community Support.}} The only statistically significant relationship we discovered in this study was between participants who included ethics in their course and those who had a professional community to discuss ethics with. This implies that one important support for educators to include ethics in their courses is to have a community that supports this endeavor. The SIGCSE community is well positioned to help push forward this goal, and is already making strides. The number of research papers published at the conference related to ethics have increased substantially since 2017 \cite{fiesler2020we}, and, subsequently, conferences have also featured relevant birds-of-a-feather and nifty assignments sessions \cite{bhalerao2022learning, doore2020assignments}, as well as workshops \cite{walsh2022making, rentz2021before}. Additionally, recent SIGCSE experience reports have introduced and demonstrated effective methods for incorporating ethics into computing courses \cite{fiesler2021integrating, KlassenFiesler2022, cs-ethics, bullock-2021}.

The importance of community also aligns with another RQ3 finding, which showed that participants need additional encouragement and support from colleagues within their institution to incorporate ethics into their courses. Top-down support for cultivating community within or between departments at a university could include hiring faculty whose primary research area is computing ethics or encouraging Sociology/Philosophy and computing professors to co-teach classes together. This could impact students in terms of expertise for standalone ethics classes, and could also help provide a resource in the form of expertise to other faculty. Another option could be to create working groups where faculty can workshop how to build ethics into existing computing courses.

\emph{\textbf{Education Advocacy \& Incentive Structures.}} Both educators who did and did not include ethics in their classrooms implicated the role of university power dynamics and labor conditions in their decision to do so. Educators who did not include ethics specifically mentioned their own lack of power to change or alter the course curriculum, while educators who did include ethics mentioned that a lack of time to make curriculum changes and a lack of institutional support on the matter made inclusion difficult. 

One approach to support this could be to revise existing incentive structures to explicitly value incorporation or creation of computing ethics content. For example, ensuring that ethics curriculum innovation is helpful to a tenure case, or providing monetary incentive, service credit or teaching releases for this kind of work. Departments could also encourage instructors to share any new ethics materials they create in public repositories such as the one introduced by \citet{fiesler2020we}. As of now, there is not incentive at an institutional level for instructors to share these course materials.

It is worth noting that 41\% of respondents were not tenure track faculty, and that these educators, such as graduate student workers and adjunct faculty, often face the most precarity in their roles. Supporting efforts around improving labor conditions for these workers, such as unionization, not only will enable these educators to advocate for more ethics in computing, it is also a way to model ethical practice for students to see. 

\emph{\textbf{When and How to Incorporate Ethics.}} When and how is it appropriate to teach students about computing ethics? About one quarter of our participants believed that students need to have a baseline level of computing skills before they can think about ethical issues in computing. Conversely, previous work has shown that students benefit when ethics is taught in introductory computing courses, because it helps them think ethically when they are learning how to code throughout the duration of their degree \cite{fiesler2021integrating}. Moreover, some participants in our survey expressed that they did not think ethics was appropriate for all computing classes, and instead, ethics belonged in a separate standalone course required by their department. However, other participants in our study reported that they would like access to modules that they could incorporate in their classes, which implied that they prefer ethics to be incorporated within existing courses throughout the entire computing degree, rather than a single standalone course. 

Previous work has recommended that in addition to incorporating ethics throughout existing computing curricula, degree programs should also include standalone classes focused specifically on computing ethics topics \cite{martin1996implementing}. This style of integration could allow for subject matter experts (including from different departments) to teach deep-dives on the theories of ethics and philosophy, while more applied and specific ethics discussions could be left to modules that match the niche topics of a computing course. To be clear, we are not suggesting that a machine learning instructor should be reading Kant in order to include ethics in their class; but as an expert on machine learning as a domain area, they should be capable of learning enough about the specific, applied area of machine learning ethics in order to competently teach that topic to students. This would also address concerns expressed by some participants about computing educator's risk of \emph{``poorly''} teaching ethics topics due to their lack of domain expertise.

\emph{\textbf{When Ethics ``Isn't Relevant.''}} Our analysis revealed that another barrier for instructors operationalizing ethics in their classroom was a feeling that ethics is not relevant to every computing topic. To address this, we suggest that future research focus on which factors contribute to educators' perceptions of relevance between any computing topic and ethics, and which interventions promote educators to see ethics as relevant or not. This barrier could also be addressed by bolstering more inter-department or interdisciplinary collaboration to create content in these ``less-applicable'' courses, or to educate computing instructors about ethics topics that could be appropriate for their course. This could also involve incorporating guest lectures from experts on niche computing ethics topics. Additionally, recent work has begun to generate coding ethics assignments for these kinds of topics, such as \citet{icer-brown}, which focuses on integrating ethics into advanced algorithms courses. Future work could pursue this kind of curriculum development, with a focus on generating ethics assignments that fit into niche or more advanced computing courses.

Given the amount of participants who shared this belief---that ethics is not relevant to their course topic---and how this feeling acted as a barrier to prioritizing ethics discussions in the classroom, we recommend that future research explores this tension.

\section{Limitations}
Our work has several limitations that are important to note. As mentioned previously, our recruitment methods may have introduced selection bias, which could limit the generalizability of our findings. Since the recruitment text included the word ``ethics,'' faculty with limited or no interest in ethics may have also declined participation and are therefore potentially underrepresented in the sample. Cross-sectional survey data limitations also include potential response biases, sample coverage, and the inability to gauge how attitudes and behaviors change over time. Additionally, since we did not provide a definition of “ethics,” this allowed for participants to respond based on their own operationalization, which may differ. However, the intent of this research was to examine perceptions towards including ethics-related content, as well as barriers and facilitating factors for increasing ethics in computing courses, not to determine differences in how ethics is defined or integrated into course content. Future studies should investigate these differences.

\section{Conclusion}
To conclude, the results of our survey showed that many computing educators see value in including ethics in computing courses, but certain barriers are preventing them from doing so. These barriers included a desire to leave ethics to other courses/departments; prioritizing technical topics over ethics topics; the feeling that ethics does not apply to some computing topics; an inability to control course content; a lack of time or incentives; and a feeling that ethics distracts from core computing topics. The structures that appeared to support ethics in the computing classroom the most were providing additional resources for educators to re-use material or make their own material from scratch; collaboration with others; and community or institutional support.

In order to incorporate more ethics into computing curricula, these existing barriers need to be addressed and current supports need to be strengthened and shared throughout the computing education system. Future research should focus on exploring the recommendations outlined in this work to more effectively support educators in incorporating ethics into their computing classrooms.

\begin{acks} 
We are grateful to Samantha Dalal, Daniella DiPaola, Stacy Doore, Julie Jarzemsky, Crystal Lee, Michael Levet, Benjamin Shapiro, Marty Wolf and members of the Internet Rules Lab for their thoughtful feedback. Special thanks to Chris Hovey and Lecia Barker for their feedback and help with email recruitment. This research was supported in part by Omidyar Network, Mozilla, Schmidt Futures and Craig Newmark Philanthropies as part of the Responsible Computer Science Challenge, and ongoing work in this space is supported by NSF IIS-2046245.
\end{acks}

\bibliographystyle{ACM-Reference-Format}
\balance
\bibliography{bibliography}










\end{document}